# Solar-driven variation in the atmosphere of Uranus

K.L. Aplin[1] and R.G. Harrison[2]

[1] Department of Physics, University of Oxford, UK

[2] Department of Meteorology, University of Reading, UK

Corresponding author: Karen Aplin (<u>karen.aplin@physics.ox.ac.uk</u>)

**Key Points**
- 11-year solar variability identified in Uranus reflectivity fluctuations
- Statistical analysis suggests fluctuations arise from solar-modulated galactic cosmic ray ions/electrons and UV-induced chemistry
- These results indicate planetary atmospheric variability can be driven by heliospheric coupling with a host star.

**Abstract**

Long-term measurements (1972-2015) of the reflectivity of Uranus at 472 and 551 nm display variability that is incompletely explained by seasonal effects. Spectral analysis shows this non-seasonal variability tracks the 11-year solar cycle. Two mechanisms could cause solar modulation, (a) nucleation onto ions or electrons created by galactic cosmic rays (GCR), or (b) UV-induced aerosol colour changes. Ion-aerosol theory is used to identify expected relationships between reflectivity fluctuations and GCR flux, tested with multiple regression and compared to the linear response predicted between reflectivity and solar UV flux. The statistics show that 24% of the variance in reflectivity fluctuations at 472 nm is explained by GCR ion-induced nucleation, compared to 22% for a UV-only mechanism. Similar GCR-related variability exists in Neptune's atmosphere, hence the effects found at Uranus provide the first example of common variability in two planetary atmospheres driven through energetic particle modulation by their host star.

**Plain Language Summary**

Measurements of the planets Uranus and Neptune have been made using a telescope, for every year from 1972 to 2015. How bright a planet appears to us is an indicator of the cloud cover in its atmosphere. An 11-year brightness variation was spotted in the Neptune observations many years ago, indicating that a process linked to the the Sun's 11-year activity cycle affects the planet's clouds. This inspired us to look at the data for Uranus more closely, and we found the same signal as for Neptune. There are two possible explanations. One possibility is chemical, when light from the sun affects the colour of particles in the planet's atmosphere. Our other possibility is that energetic particles from outside the Solar System, cosmic rays, influence particle or cloud formation. (Cosmic rays are "bent" away from the Solar System by the Sun acting as a magnet, so are also affected by its 11-year activity cycle). In our results, we actually find that both of them have a small effect on the clouds on Uranus. This is the first evidence of two planetary atmospheres – Neptune originally and now Uranus - showing similar variations, in both cases originating from their host star.

## 1. Introduction

Lockwood and Jerzykiewicz (2006) have been routinely measuring the reflectivity of Neptune and Uranus from Earth for over 40 years, effectively providing a long-term record of disk-averaged cloud cover. A solar cycle in the Neptune data, and therefore in the atmosphere of Neptune, was first identified by Lockwood and Thompson (1986). It remained unexplained until Aplin and Harrison (2016) demonstrated that the reflectivity fluctuations were associated with both galactic cosmic rays (GCR) and ultraviolet (UV) radiation modulating Neptune's clouds through ion-induced nucleation (Moses et al, 1992) and UV-induced colour changes (Baines and Smith, 1990) respectively. The ubiquity of atmospheric ionization from GCR (e.g. Aplin, 2006), and the known similarities between the atmospheres of Neptune and Uranus (e.g. Lunine, 1993), motivate re-examination of the Uranus reflectivity data.

A time series of the Uranus reflectivity measurements is shown in Figure 1a. Measurements were made from 1972-2015 with a 21-inch telescope at Lowell Observatory, Arizona, and are carefully calibrated against stars of known reflectivity (Lockwood and Jerzykiewicz, 2006).  Two filters in the visible region (*b* (442nm) and *y*

(551nm)) were used to measure the reflectivity of Uranus on several nights every year when it is closest to Earth. There is a dominant periodicity of approximately 42 years, which is seasonal (e.g. Miner, 1998). (For comparison, one year on Uranus is 84 Earth years). Hammel and Lockwood (2007) predicted that the planet would begin to brighten in 2006 because of seasonality i.e. geometrical variations due to the parts of the planet seen from Earth, as in figure 1a. However, as for Neptune, seasonal effects do not fully explain the observed variability in reflectivity (Hammel and Lockwood, 2007). To investigate the remaining variability, a robust non-parametric local smoother, weighted towards points near the region to be fitted, known as LOESS (LOcally Estimated Scatterplot Smoothing) (Cleveland et al, 1992) was applied, avoiding the need to assume a model for the planet's seasonal reflectivity variations as previously (Hammel and Lockwood, 2007), Figure 1a. (A similar detrending technique, LOWESS (LOcally WEighted Scatter-plot Smoother) (Cleveland et al, 1992), with different smoothing characteristics, was also investigated but did not follow the observations as effectively as LOESS. This contrasts with the analysis carried out for Neptune by Aplin and Harrison (2016) where the LOESS and LOWESS detrends were indistinguishable.) The detrended data, i.e. raw data with the LOESS fit subtracted, is shown in Figure 1b, with the solar-modulated parameters UV and GCR plotted as Figures 1c and 1d (data sets are described in section 3.1). Figures 1b, 1c and 1d indicate common variations in the reflectivity fluctuations and solar UV radiation, and an antiphase relationship between Uranus's reflectivity and GCR. This suggestive result inspired further data analysis, firstly calculation of power spectra to search for a solar periodicity signal, and secondly, multiple regression to assess the contribution of the two solar-modulated mechanisms, UV and GCR.

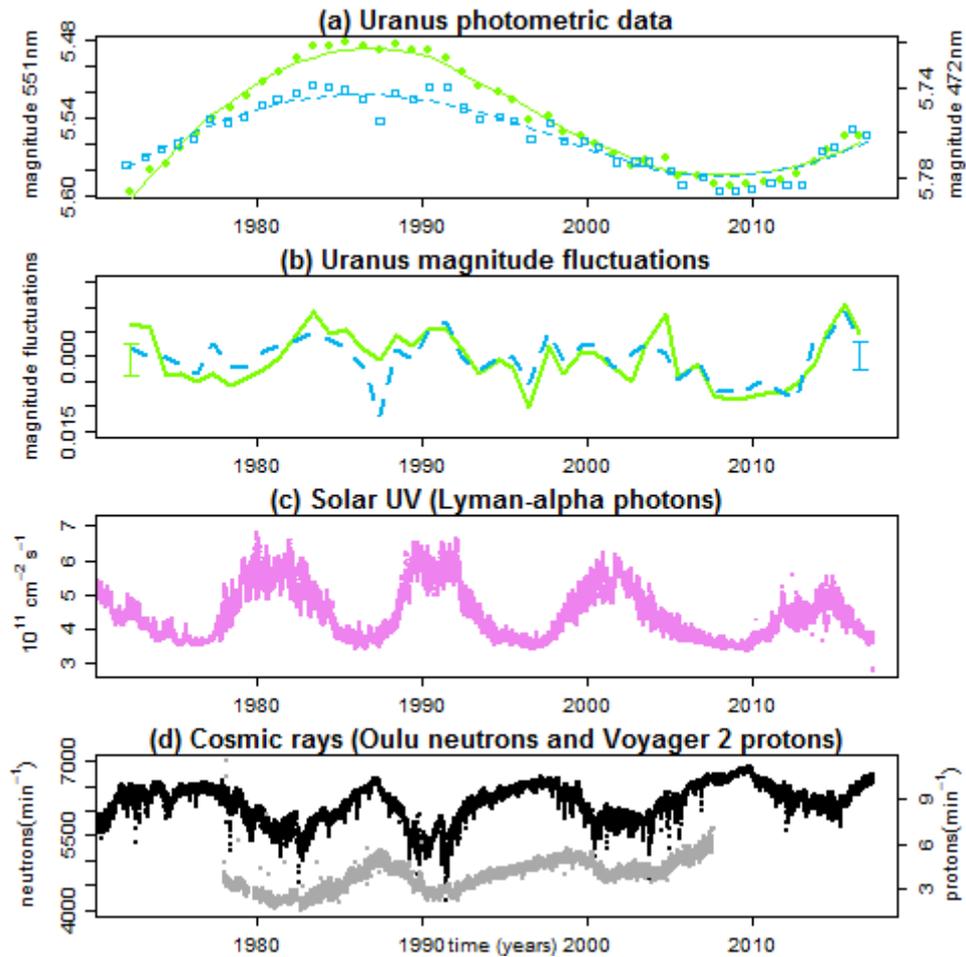

Figure 1. Time series of Uranus's brightness, solar ultra-violet (UV) radiation and galactic cosmic rays (GCR). (a) Uranus's brightness (astronomical magnitude, where smaller values represent a greater signal) time series at 472nm (blue squares) and 551nm (green circles), each smoothed with a LOESS fit (blue dashed line or green solid line). Data is from Lockwood (2017) and is described fully in Lockwood and Jerzykiewicz (2006). (b) Magnitude fluctuations after detrending (a) with a LOESS fit, weighted by the standard error in each measurement. Typical 95% confidence limits on the mean for each point, obtained from the standard error on the LOESS fit, are shown as a single error bars on the far left and far right. (c) Lyman alpha (UV) radiation at 121.5nm ($10^{11}$ photons cm$^{-2}$ s$^{-1}$), from LISIRD (2017). (d) GCR at Earth's surface and in the heliosphere, showing terrestrial neutron monitor data from Oulu, Finland, in daily averaged counts min$^{-1}$ (black) (Usoskin, 2017) and Voyager 2 Low Energy Charged Particle (LECP) instrument daily mean flux of GCR protons > 70MeV (grey) in min$^{-1}$ (Decker et al, 2015). Standard deviations are typically 4% in the LECP data.

## 2. Spectral analysis

Power spectra of the detrended data (Figure 1b) were made using the Lomb-Scargle algorithm (Lomb, 1976; Scargle, 1982) implemented in R (Press and Rybicki, 1989), with a cosine bell taper parameter, applied to the entire data set, of 0.1. The results, Figure 2, show a statistically significant 11-year solar cycle in both the filters, which are robust to the errors in the measurements (Lockwood, 2017). Other periodicities are present, for example, at ~8 years in Figure 2a, similar to those previously reported by Ashbrook (1948) and Hollis (2000), and at 16.5 years in Figure 2b. The 8, 11 and 16.5 year periodicities remain if the LOWESS detrending technique (Cleveland et al, 1992) is used, so are unlikely to be an artefact of the detrending. They are also numerically unrelated to the 42 year seasonality. The periodicities at 8 and 16.5 years currently lack any empirical or theoretical explanation, but imply a range of origins for atmospheric modulation.

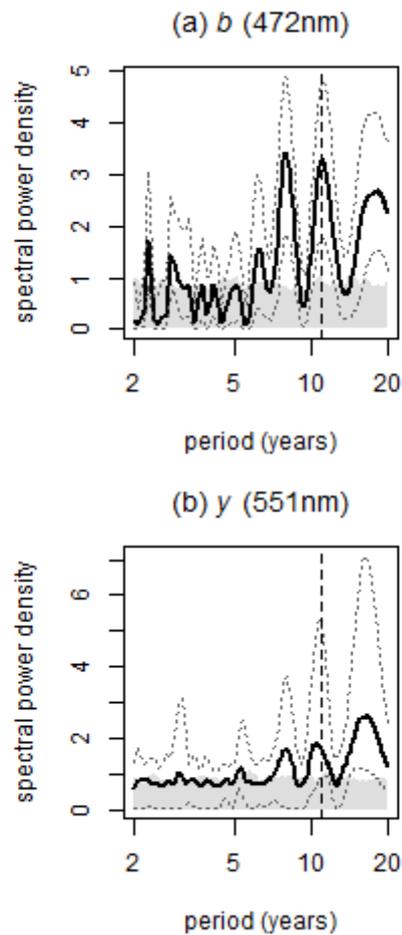

Figure 2. Dimensionless power spectral density (thick black lines) calculated for (a) 472 nm and (b) 551 nm Uranus magnitude data using the Lomb-Scargle periodogram method after de-trending with a LOESS fit. The 11-year solar cycle is indicated as a dashed vertical line. The statistical significance of the spectral peaks was estimated using Monte-Carlo procedures. The grey shading shows the positive confidence range (i.e. up to the median signal plus the 95th percentile) of 10000 realisations of the power spectra calculated in the same way as the spectral peak, but after random shuffling of the magnitude data. The thin dashed lines mark 95% confidence limits from 10000 realisations of the power spectra calculated with the uncertainties in the magnitude fluctuations quoted by Lockwood (2017).

Two suggested mechanisms thcould cause a solar cycle variation in the atmosphere of Uranus, expressed through reflectivity variations. Firstly, cloud formation through condensation of supersaturated vapour (probably methane or butadiyne (diacetylene)) onto ions or electrons created by GCR (Moses et al, 1992). This mechanism was originally suggested for Neptune, but since the cloud, aerosol and atmospheric structures of Uranus and Neptune are similar (Mousis et al, 2017), the same mechanisms should be considered for Uranus. Baines and Smith (1990) proposed a mechanism to explain the Neptune solar cycle variation by which aerosols were photochemically "tanned", changing the planetary albedo; this could potentially also act on Uranus. According to Sromovsky et al (2011), the optical depths at 551nm and 472nm reach unity at atmospheric pressures of 3.5 and 2 bar respectively. This suggests that both wavelengths provide information from the troposphere, near methane and hydrogen sulphide cloud layers (Mousis et al, 2017).

The UV and GCR effects are not mutually exclusive and, as for Neptune (Aplin and Harrison, 2016), may act in combination to contribute to the observed disk-averaged reflectivity variations. In the next section, multiple regression is used to investigate the contribution of the two suggested mechanisms to the solar cycle signal apparent in the spectral data.

### 3. Statistical analysis based on ion-aerosol theory

The relationship between UV and observed reflectivity changes is expected to be linear (Baines and Smith, 1990), and GCR effects are expected to follow ion-aerosol theory. For the latter case, if ions and electrons created by GCR influence clouds, then the planet's disk-averaged reflectivity should be related to the ion and electron number concentration $n$, which are assumed to be equal (Aplin and Harrison, 2016). GCR is the only significant source of atmospheric ionisation expected in the stratosphere and troposphere of Uranus (Aplin, 2006), so the rate of change of ion/electron concentration $dn/dt$ is linked to the GCR ion production rate $q$ through the ion balance equation (1) below (e.g. Harrison and Carslaw, 2002). Loss terms due to recombination and attachment are quantified, respectively, by a recombination coefficient $\alpha$, and an attachment coefficient $\beta$ to monodisperse aerosol particles with number concentration $Z$.

$$\frac{dn}{dt} = q - \alpha n^2 - \beta n Z \ (1)$$

The simplest way to investigate whether planetary reflectivity is associated with the ionisation rate is to consider the relationship between $q$ and $n$, following two limiting cases of equation (1). When there are few pre-existing aerosol particles, $n$ is controlled by self-recombination and the aerosol term can be neglected, hence $n \propto \sqrt{q}$. In the case of a substantial background aerosol concentration, the ion concentration is limited by attachment to aerosol, with recombination ignored, so that $n \propto q$. Aplin and Harrison (2016) considered all statistical relationships within the parameter space of equation (1) and its limiting cases, and found that including both UV and GCR ionisation improved the fit over UV radiation alone. This approach is now developed to include the possible effect of ion-induced nucleation, which is used to guide the statistical modelling. Equation (1) is modified to include an ion-induced nucleation term $\gamma$, proportional to ion/electron concentration $n$ (Harrison, 2000)

$$\frac{dn}{dt} = q - \alpha n^2 - \beta n Z - \gamma n \ (2)$$

Here a simple limiting approach is taken of assuming that some, all or none of the aerosol in the atmosphere of Uranus is created by ion-induced nucleation. No assumptions are made about the location, type, or distribution of atmospheric aerosol, beyond the idea that it must exist close to the region of the atmosphere where the $b$ and $y$ filter reflectivity signals originate. These assumptions are used to define a set of physically plausible relationships for Uranus, beyond those selected by Aplin and Harrison (2016) for Neptune.

If all the atmospheric aerosol is produced by ion-induced nucleation, it is assumed that
$$Z = \gamma n \ (3).$$
Substituting eq (3) for $Z$ in equation (2) and rearranging gives
$$\frac{dn}{dt} = q - (\alpha + \beta\gamma)n^2 - \gamma n \ (4).$$

For steady state, i.e. $\frac{dn}{dt} = 0$, then the quadratic solution formula can be used to show that $n$ is expected to scale with $\sqrt{q}$:

$$n = \frac{\gamma \pm \sqrt{\{\gamma^2 + 4q(\alpha + \beta\gamma)\}}}{-2(\alpha + \beta\gamma)} \ (5),$$

corresponding to case A in table 1.

Another – and probably more likely - possibility is that some of the aerosol is produced by ion-induced nucleation and some modulated by UV-induced alterations. If two terms are used to represent the aerosol ($Z_p$ for photochemical aerosol and $Z_i$ for ion-induced aerosol), then equation (1) can be written as:

$$\frac{dn}{dt} = q - \alpha n^2 - \beta n(Z_p + Z_i) \ (6).$$

Substituting for $Z_i$ with eq (3) and rearranging gives a quadratic in $n$,

$$\frac{dn}{dt} = q - (\alpha + \beta\gamma)n^2 - \beta Z_p n \ (7).$$

If steady state is again assumed, the quadratic solution formula can be used to estimate the variation of $n$ with $q$, which gives $n$ proportional to $\sqrt{q}$:

$$n = \frac{\beta Z_p \pm \sqrt{\{(\beta Z_p)^2 + 4q(\alpha + \beta\gamma)\}}}{-2(\alpha + \beta\gamma)} \ (8),$$

corresponding to case B in table 1. Taken together, equations (5) and (8) show that if ion-induced nucleation is involved at all in aerosol production at Uranus, then the associated reflectivity variations are expected to be proportional to $\sqrt{q}$ (cases A and B in Table 1).

If there is no ion-induced aerosol particle production, the planetary reflectivity fluctuations should be proportional to the UV flux. For low concentrations of aerosol, case C in Table 1 can be assumed. However, in cloudy or aerosol-rich parts of the atmosphere, the ion balance equation (Equation 1) will be in the attachment limit, with all ions attaching to aerosol. In this case, the reflectivity fluctuations are expected to be proportional to $n$ and therefore to $q$, as well as the UV flux (case D in Table 1).

| Case | Process | Expected relationship with brightness fluctuation |
|---|---|---|
| A | All aerosol produced by ion-induced nucleation | $\sqrt{GCR}$ |
| B | Some aerosol produced by ion-induced nucleation | $\sqrt{GCR}$ |
| C | All aerosol is produced photochemically (low concentrations; recombination limit in ion-aerosol theory) | UV |
| D | All aerosol is produced photochemically (cloudy/aerosol-rich regions; attachment limit in ion-aerosol theory) | UV + GCR |

Table 1 Expected relationships, based on ion-aerosol theory, between brightness fluctuations, galactic cosmic rays (GCR) and UV flux for different solar-modulated aerosol generation processes in the atmosphere of Uranus.

Following Aplin and Harrison (2016), a sub-set of possible statistical relationships between the measured reflectivity fluctuations $f$, the UV and GCR fluxes were investigated based on equation (9) where $f_{b,y}$ are the measured magnitude fluctuations in the $b$ (472nm) or $y$ (551nm) wavelength ranges, $\kappa$, $\lambda$ and $\mu$ are coefficients for the $b$ or $y$ data representing the UV mechanism, ion attachment and ion recombination respectively

$$f_{b,y} = \kappa_{b,y}UV + \lambda_{b,y}GCR + \mu_{b,y}\sqrt{GCR} \qquad (9).$$

The initial assumption that some, all or none of the aerosol in Uranus's atmosphere is created by ion-induced nucleation, ultimately leads to three physical models. Equations (5) and (8) show that if some or all of the aerosol is from ion-induced nucleation, the reflectivity will depend on the square root of the GCR ionisation rate. If the aerosol is entirely UV-modulated, with no role for ion-induced nucleation, two cases are considered. In clear air, a UV-only reflectivity dependence is expected, but if the UV modulation occurs in cloud or other regions of high aerosol, an additional term accounts for attachment of the UV-modulated aerosol to ions or electrons. The three distinct physical models derived from the theoretical considerations outlined above are summarised in table 2. Before discussing the results, the data sets used will first be described.

| Case in table 1 | Physical interpretation | Coefficients in equation (7) | Adjusted coefficient of determination ($R^2$) *and statistical significance p* | |
|---|---|---|---|---|
| | | | *y* (551nm) | *b* (472nm) |
| A/B | √GCR only (some or all aerosol is produced by ion-induced nucleation) | $\kappa$=0, $\lambda$=0; $\mu$ free to vary | 0.17 *(p<0.005)* | 0.24 *(p<0.0005)* |
| C | UV only; low aerosol concentrations | $\lambda$=0, $\mu$=0; $\kappa$ free to vary | 0.11 *(p<0.02)* | 0.19 *(p<0.005)* |
| D | UV only; cloudy / aerosol-rich regions with ion-aerosol attachment | $\mu$=0; $\kappa$ and $\lambda$ free to vary | 0.15 *(p<0.002)* | 0.22 *(p<0.005)* |

Table 2 Summary of multiple regression analysis. Fits are weighted according to the errors on the measurements. Statistical significances (*p*-values) of the fits are indicated. The adjusted coefficient of determination ($R^2$) gives the fraction of the variance explained by the fit, whilst accounting for the different number of variables in each fit.

## 3.1 Cosmic ray and UV data

The UV dataset used is the well-known composite Lyman-alpha series of the solar hydrogen 121.57 nm emission line (LISIRD Data Systems Group, 2017), estimated to be accurate to ± 10%. There are no GCR data at Uranus spanning the 45 years of its reflectivity observations, so terrestrial neutron monitor data from the Oulu station in Finland have been used (Usoskin, 2017). Neutrons are secondary GCR particles generated by the atmospheric decay of GCR, and are a reasonable proxy for ionisation in the whole atmosphere column (Harrison et al, 2014). Although there is some *in situ* GCR data from Voyager 2's Low Energy Charged Particle (LECP) detector, which passed Uranus in 1986 (Decker et al, 2015), it is preferable to use terrestrial GCR measurements as an indicator of ionisation at Uranus for three reasons. Firstly, GCR originate from energetic events well beyond the Solar System. There is some heliospheric modulation of lower-energy particles, but the flux of primary energetic particles (the secondaries from which are mainly responsible for tropospheric ionisation) can be assumed to be essentially constant (Moral, 2014; Usoskin and Kovaltsov, 2006), justifying the use of an Earth-based counter. (Figure 1b shows that the energetic particles measured by Voyager 2 varied more with the solar cycle on its journey through the Solar System than with spacecraft distance from the Sun, which was increasing throughout the time series). Secondly, there is a small but variable lag in the GCR measurements (of up to 4 months) between Voyager 2 and Earth. This lag between space and earth-based GCR data is well-known and arises from several effects, some of which (such as changes in the size of the GCR modulating region) could be related to the spacecraft position (e.g. Van Hollebeke et al, 1972). Because of this, a simpler approach is to use data from a fixed location for analysis (Aplin and Harrison, 2016). Finally, the Oulu station is long-established and the measurements are essentially unchanged since the 1960s, whereas Voyager 2 only measured heliospheric GCR between 1977 and 2007 (Decker et al, 2015).

As in Aplin and Harrison (2016), 20-day averages of the UV and GCR data centred on the observation dates given in Lockwood (2017) were calculated to use in the multiple regression. Regressions were weighted according to the measurement error

(Lockwood, 2017), with UV and GCR errors considered negligible with respect to the uncertainty in the telescope data.

## 3.2 Results

The results of the regression analysis are shown in Table 2. The fits for Case A/B (ion-induced nucleation from GCR ionisation) and D (UV "tanning" in the presence of background aerosol) are similar in terms of the amount of variance in the Uranus reflectivity variations they explain. The fit for Case A/B is plotted in Figure 3, and the other fits in the Supplementary Figures.

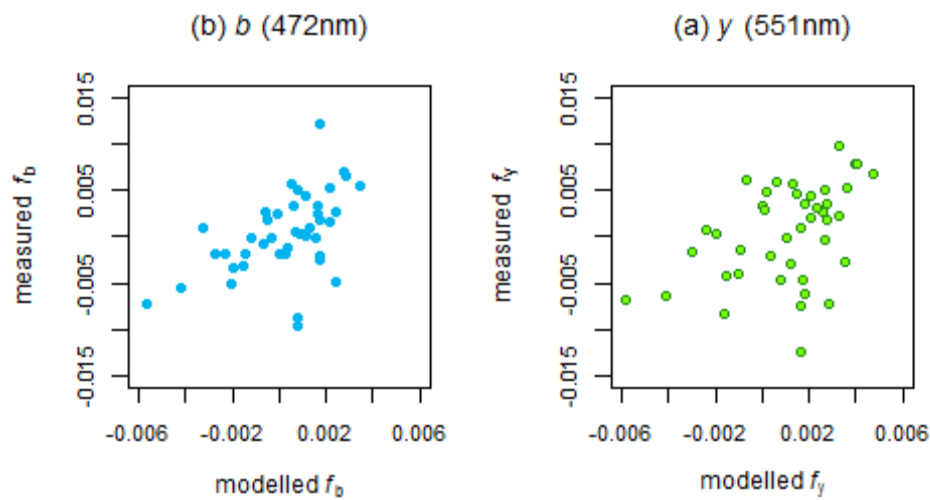

Figure 3 Physically realistic linear regression models to explain Uranus magnitude fluctuations. The fit for a model including ion-induced nucleation, based on a square root relationship between the brightness fluctuations and the cosmic ray flux (Case A/B in table 1) is shown in each case from the range of possible relationships tested.

To aid comparisons of the magnitudes of the proposed effects between Uranus and Neptune, normalised coefficients were generated by calculating the fits with respect to the averaged GCR and UV variations for each data set (Table 3). At Neptune (for which a version of Table 3 is given in the Supplementary Information), the best statistical relationship implies a UV-dominated mechanism over the time series, with a comparable sensitivity to UV across both wavelengths at about 0.03 mag/fractional change in Lyman-alpha flux, where mag indicates the change in astronomical magnitude. At Uranus, the ion-induced nucleation effect was significantly greater than for Neptune (see Supplementary Table 1), at 0.06 or 0.07 mag/fractional change in GCR neutron flux.

|  |  | **Coefficients from equation (1) with units and (physical interpretation)** |  |  |  |
| --- | --- | --- | --- | --- | --- |
| **Wavelength** | **Model (#in table 1)** | $\kappa$ **(UV)** | $\lambda$ **(GCR)** | $\mu$ **($\sqrt{}$ GCR)** | **adjusted R$^2$ ($p$-value)** |
| 472 nm | $\sqrt{}$GCR (A/B) | NA | NA | 0.06±0.02 | 0.24 (*p<0.0005*) |
| 551 nm |  | NA | NA | 0.07±0.02 | 0.17 (*p<0.005*) |
| 472 nm | UV (C) | -0.011±0.003 | NA | NA | 0.19 (*p<0.005*) |
| 551 nm |  | -0.011±0.004 | NA | NA | 0.11 (*p<0.02*) |
| 472 nm | UV + GCR (D) | -0.001±0.006 | 0.03±0.02 | NA | 0.22 (*p<0.005*) |
| 551 nm |  | 0.002±0.009 | 0.04±0.02 | NA | 0.15 (*p<0.02*) |

Table 3 Normalised coefficients for fits to each of the proposed models for 472 nm and 551 nm data at Uranus. NA for Not Applicable indicates that a coefficient was not relevant for that fit. Errors are the standard error in the fit. Adjusted coefficient of determination and statistical significance (*p*-value) are also shown.

## 4. Discussion

An 11-year periodicity is present in fluctuations in Uranus's reflectivity from 1972-2015, with possible solar-driven coupling mechanisms being through UV or GCR ion-induced nucleation. Statistical modelling based on ion-aerosol theory, initially presented by Aplin and Harrison (2016) has been developed further here to distinguish the solar-modulated physical mechanisms, and indicate that GCR ion-induced nucleation (Case A/B) and a UV effect (Case D) are indistinguishable, explaining ~20 % of the variance in Uranus's reflectivity variations. This provides evidence for both ion-induced nucleation and UV effects in the troposphere. The improved statistics of Case D over Case C imply that the UV effect is taking place in a region of high background aerosol concentrations, perhaps in a cloud layer. The statistical evidence is also consistently stronger for 472nm (*b*) over 551nm (*y*), indicating that both the UV and GCR effects are occurring nearer 2 bar than 3.5 bar. It should be noted that both the mechanisms described only account for one fifth of the variance in the data, indicating that other effects are likely to dominate variability in Uranus's atmospheric reflectivity. This is supported by the unexplained spectral peaks at 8 years and 16.5 years in Figure 2.

There are no studies of ion-induced nucleation on Uranus, but modelling of Neptune suggested that both methane and diacetylene (butadiyne) in the troposphere and stratosphere respectively may become sufficiently supersaturated for ion-induced nucleation (Moses et al, 1992; Aplin and Harrison, 2016). On Uranus, the methane cloud layer, also at about 1 bar and 75K, may support ion-induced nucleation if there is suitable supersaturation (Mousis et al, 2017; figure 7a in Aplin and Harrison, 2016). The $b$ and $y$ filters could correspond to the hydrogen sulphide cloud layer, but a lack of data on the physical properties of $H_2S$ at low temperature makes it difficult to estimate whether ion-induced nucleation is likely.

The $b$ and $y$ filter data analysed here are unlikely to be sensitive to the thin stratospheric diacetylene haze layer. However, the possibility of ion-induced nucleation in the stratosphere of Uranus is implied by a pre-Voyager study (Atreya et al, 1991), arguing that conservation of mass forces an increase in diacetylene supersaturation of eight orders of magnitude within the haze layer. This means that the supersaturation at which it becomes energetically favourable for gas to condense onto an ion is more likely to be achieved (e.g. Mason, 1971). If the stratospheric diacetylene layers are at similar temperatures and pressures on Uranus as Neptune, but more highly supersaturated, then ion-induced-nucleation would be favoured on Uranus (see figure 7b in Aplin and Harrison, 2016). Further work is needed to investigate ion-induced nucleation in this region of the atmosphere, and the possibility that aerosols formed in the stratosphere could perhaps also modulate tropospheric cloud reflectivity through sedimentation (e.g. Lunine, 1993).

These findings both identify a solar cycle signal in the atmosphere of Uranus for the first time, and indicate a clear role for ion-induced nucleation and UV in non-seasonal reflectivity fluctuations from 1972-2015. This is the first evidence for the existence of common variations in planetary atmospheres due to energetic particle modulation by the parent star.

## Acknowledgements

All data for this paper is cited and listed in the reference list. The uniquely valuable measurements of Dr W. Lockwood over many years are particularly acknowledged. Dr L. Fletcher and Dr C. Arridge contributed considerably to the development of this manuscript through helpful and constructive reviews.